\documentclass[aps,preprint]{revtex4}
\usepackage{graphicx,color}

\begin{document}
\title[Internal-tide scattering]{A laboratory study of low-mode internal tide scattering by supercritical topography}

\author{Thomas Peacock$^{1}$, Matthieu Mercier$^{2}$, Henri Didelle$^{3}$, Samuel Viboud$^{3}$, Thierry Dauxois$^{2}$}
\date{\today}
\affiliation{1. Department of Mechanical Engineering,
Massachusetts Institute of Technology, Cambridge, MA 02139, USA\\
2. Laboratoire de Physique de l'\'Ecole Normale Sup\'erieure de
Lyon, Universit\'e de Lyon, CNRS, 46 All\'{e}e d'Italie, 69364 Lyon
c\'{e}dex 07, France\email{Matthieu.Mercier@ens-lyon.fr,Thierry.Dauxois@ens-lyon.fr}\\
3. Coriolis Platform, Laboratoire des \'Ecoulements G\'eophysiques
et Industriels, CNRS-UJF-INPG, 21 rue des Martyrs, 38000 Grenoble,
France\\}

\bibliographystyle{unsrt}

\begin{abstract}
We present the first laboratory experimental results concerning the scattering of a low-mode internal tide by topography. Experiments performed at the Coriolis Platform in Grenoble used a recently-conceived internal wave generator as a means of producing a high-quality mode-1 wave field. The evolution of the wave field in the absence and presence of a supercritical Gaussian was studied by performing spatiotemporal modal decompositions of velocity field data obtained using Particle Image Velocimetry (PIV). The results support predictions that large-amplitude supercritical topography produces significant reflection of the internal tide and transfer of energy from low to high modes.
\end{abstract}
\maketitle


The generation of internal tides, which are internal wave fields of tidal period, has been the subject of much recent study~\cite{Garrett}. A combination of satellite altimetry, field studies, analysis, numerics and laboratory experiments, estimate there to be an energy flux into the internal tide of around 1TW . It is believed that this energy flux plays a crucial role in ocean mixing, although its ultimate fate remains to be clearly established.

Several examples of mechanisms leading to dissipation of the internal tide are currently being investigated. Internal tides of short vertical wavelength (i.e. high modes) are expected to be unstable near their generation site, where they can immediately produce overturnings and mixing~\cite{Lueck}. Most of the internal tide energy, however, is believed to be radiated with long vertical wavelength (i.e. as low modes) away from the generation site~\cite{Ray,Echeverri}, to participate in wave-wave interactions~\cite{Alford}, (critical) reflection at sloping boundaries~\cite{Dauxois}, and scattering by mesoscale structures~\cite{Rainville} and finite-amplitude bathymetry~\cite{Johnston2}. A long-term goal is to determine the relative importance of these processes in the dissipation of the internal tide.

There have been a number of two-dimensional, linear analytical studies of internal tide scattering by subcritical topography, where the slope of the topography is small compared to the ray slope of internal waves~\cite{Baines1,Baines2, Muller, StLaurent}; in all these cases the ocean was treated as being of infinite depth. A general conclusion of this body of work is that there is a redistribution of internal tide energy flux from low to high mode numbers (i.e from long to short vertical wavelengths), which increases the likelihood of breaking and mixing. This redistribution seems to be around 6-10\% of the incoming energy
flux, suggesting that scattering of the low-mode internal tide by sea floor topography is not a dominant process in the ocean. There is some ambiguity, however, as it is found that scattering becomes more efficient for larger and more rugged topography, where small-amplitude theories ultimately break down, in which case it could be an $O$(1) process for the deep ocean \cite{Muller}.

Studies of finite-amplitude, two-dimensional topography in a finite-depth ocean reveal scenarios where internal tide scattering by bathymetry is a dominant process. For example, analytical studies of a knife-edge barrier \cite{Larsen,Robinson}, and slope-shelf and ridge configurations \cite{MullerLiu1,MullerLiu2} reveal that substantial scattering can occur; the extent of scattering depending on the criticality of the topography and the depth-ratio. A general result is that large amplitude subcritical and supercritical topography transmit and reflect the internal tide, respectively.

The impact of scattering by finite-amplitude bathymetry has also been revealed by numerical simulations of idealized two- and three-dimensional topographic features, and by processing of satellite altimetry data from the Line Islands ridge located south of Hawaii \cite{Johnston1, Johnston2}. These studies report up to 40\% of the incident energy flux of a mode-1 internal tide being scattered, and a corresponding enhancement of mode-2, leading to the conclusion that scattering may be important in topographically rough regions of the Pacific.

Although there have been studies of the interaction of internal tide beams with sloping
boundaries~\cite{Ivey, Peacock, Dauxois} and tidal interactions with continental shelves~\cite{Ivey2}, to the best of our knowledge there are no reported laboratory experimental results of low-mode internal tide scattering by bathymetry. A principal reason is that it has not previously been feasible to produce a low-mode internal tide whose evolution can be studied in a controlled, experimental setting. In this Letter, we demonstrate a new method of reliably generating high-quality low-mode internal tides using a recently conceived internal wave generator~\cite{Gostiaux, Mercieretal}. In turn, this allows us to perform the first experimental studies of low-mode internal tide scattering by bathymetry, utilizing a modal decomposition algorithm that has been previously employed for experimental studies of internal tide generation~\cite{Echeverri}. This breakthrough sets the scene for future, systematic experimental studies of internal tide scattering, to complement ongoing theoretical and numerical studies, and upcoming field programs.

\begin{figure}
\begin{center}
\includegraphics[height=7cm]{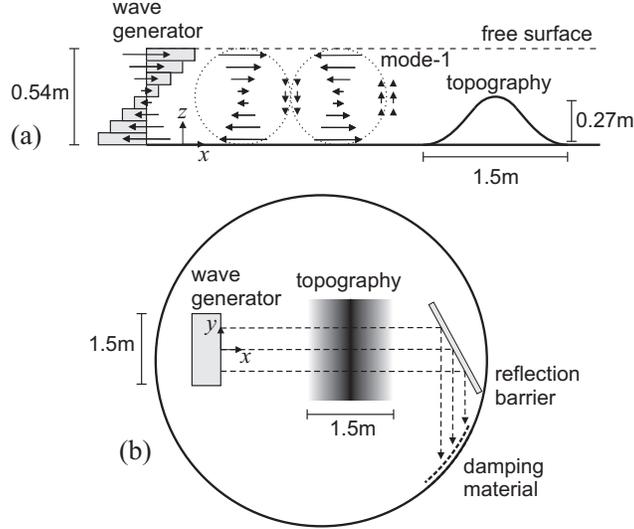}
\end{center}
\caption{A schematic of the experimental configuration as viewed from (a) the side and (b) above.
The plates of the internal-wave generator are configured to match the horizontal velocity component of a mode-1
internal tide. The radiated internal tide is incident on the Gaussian topography and the transmitted wave field is ultimately
directed away from the interrogation region by a reflection barrier, after which the wave field is also damped.}\label{FigSchematic}
\end{figure}

Experiments were performed in the 13~m diameter circular tank of the Coriolis Platform in Grenoble; a schematic of the experimental arrangement is presented in figure~\ref{FigSchematic}. The tank was filled to a depth of $H=0.54$ m with a near-linear stratification of characteristic Brunt-V\"ais\"al\"a frequency $N=0.63\pm 0.07$ rad/s (i.e. period $\tau\approx\/10$~s), as determined using a calibrated PME conductivity probe. Slight variations of the stratification away from linearity were primarily due to the diffusive evolution of the stratification near the top and bottom boundaries, and are accounted for in our analysis.

A mode-1 internal tide was generated using the $1.5$~m wide, $0.54$~m tall internal-wave generator previously employed to study plane wave reflection from a sloping boundary~\cite{Gostiaux}. The generator comprised 24 horizontally-moving plates that were
configured for these experiments as illustrated by the schematic in figure~\ref{FigSchematic}(a), with maximum,
anti-phase displacements at the bottom and top to mimic the horizontal velocity component of the mode-1 internal tide in a linear stratification. More precisely, the amplitude of displacement of the plates was set to $A_0\cos(\pi z/H)$, with $A_0=35\pm1$ mm and the bottom and top of the generator corresponding to $z=0$ and $H$, respectively. The frequency of oscillation for the experimental results presented here was $\omega=0.35\pm0.005$ rad/s.

The horizontal distance between the internal-wave generator and the far wall of the experimental tank was $8$~m, and we define $x$ as the corresponding horizontal coordinate originating at the generator (see figure~\ref{FigSchematic}). An angled reflection barrier was installed at the far wall to prevent internal waves being reflected back into the field of view.
As a precaution, damping material was also placed in the path of waves reflected away by the barrier, to further reduce the possibility of them having any subsequent impact. This configuration proved highly successful, as the experiments were run for up to ten minutes ($> 30$ tidal periods) without any evidence of back-reflected waves in the field of view.

The topography used to scatter the incident mode-1 internal tide was a nominally two-dimensional Gaussian ridge of cross-section $h(x)=h_0\exp(-x^2/2\sigma^2)$ constructed from PVC. The principal dimensions of the ridge were height $h_0$ = 0.273 m and variance $\sigma$ = 0.3 m, so that the the ridge extended for roughly 1.5 m in the $x$-direction and had a depth ratio $h_0/H$=0.5. In the  transverse, $y$-direction the span of the topography was $\ell=2.0$~m, which was 0.5~m greater than the span of the generator. The first set of experiments were conducted without the topography present, to investigate the integrity of the mode-1 tide. Thereafter, the ridge was introduced with its center at $x=3.93$~m, which was sufficiently far to ensure that a clean mode-1 wave field was established by the time it reached the topography. There was no discernable effect due to internal tides being reflected back and forth in the region between the generator and the topography.

Particle Image Velocimetry (PIV) was used to measure the velocity field. Snapshots of the flow were recorded at $3$ or $4$~Hz using a pair of 12-bit CCD cameras with a resolution of 1024 $\times$ 1024 pixels. The Correlation Image Velocimetry algorithms available through the Coriolis facility~\cite{Fincham} were used to calculate the velocity field from the experimental movies. Finally, the Hilbert transform technique~\cite{Garnier} was employed to
distinguish the fundamental frequency component of the wave field from any higher harmonics produced by nonlinearity, and to confirm the absence of back-reflected waves.

\begin{figure}
\begin{center}
\includegraphics[width=0.95\linewidth]{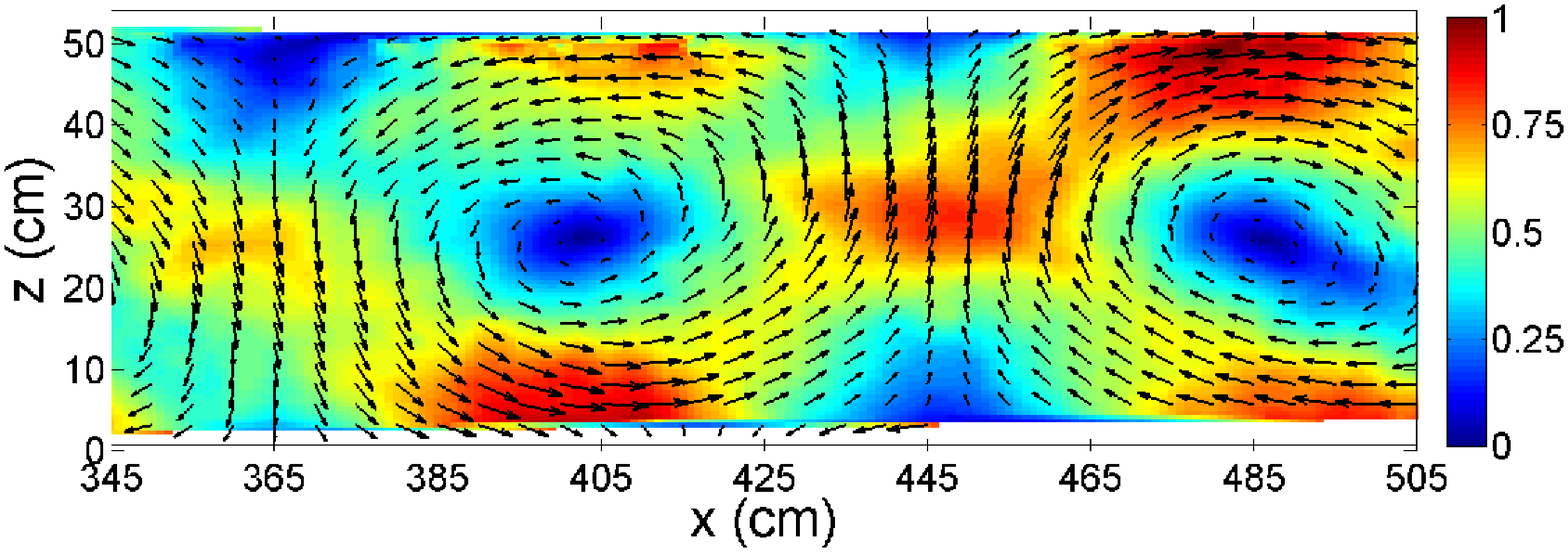}\\
\includegraphics[height=4cm]{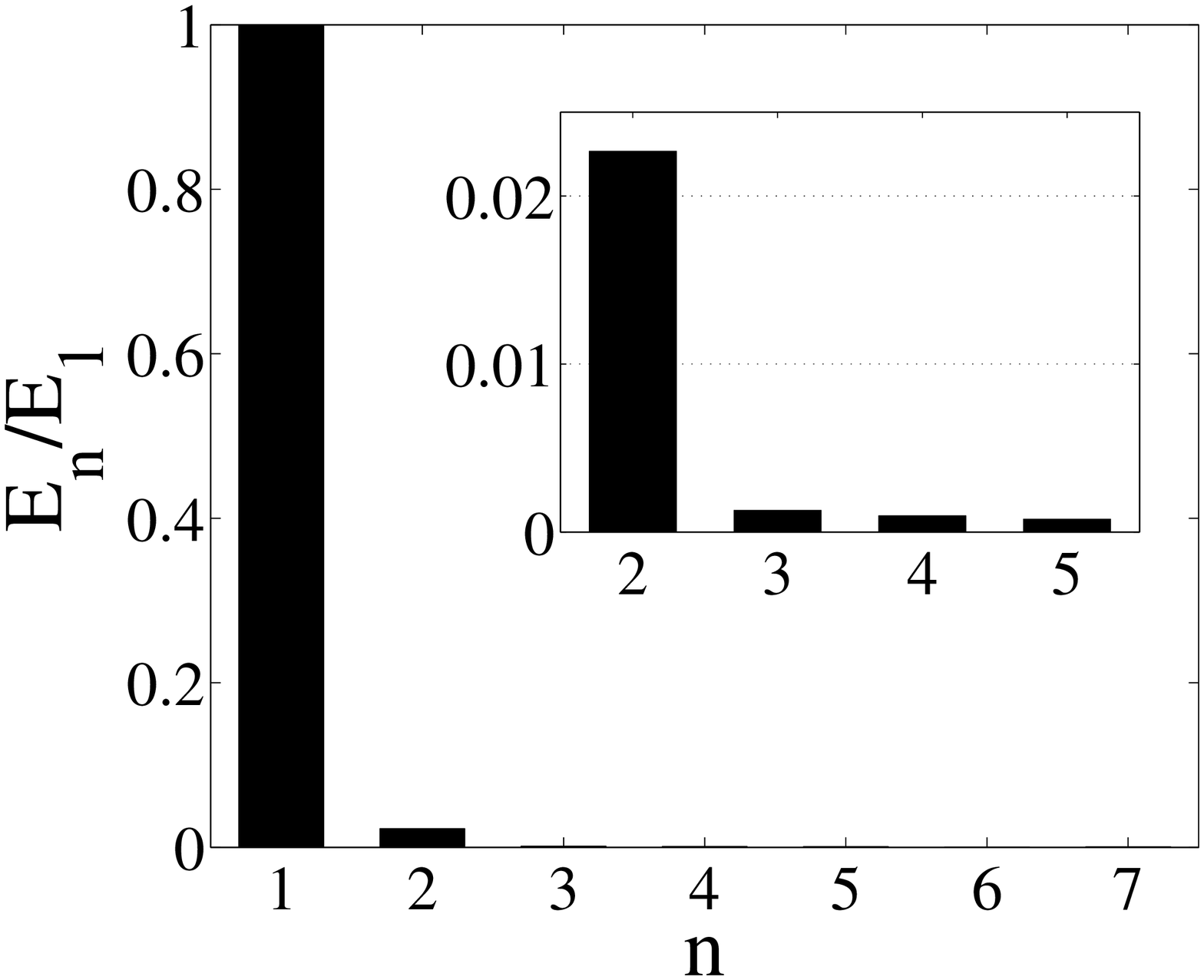} \hspace{2cm}
\includegraphics[height=4cm]{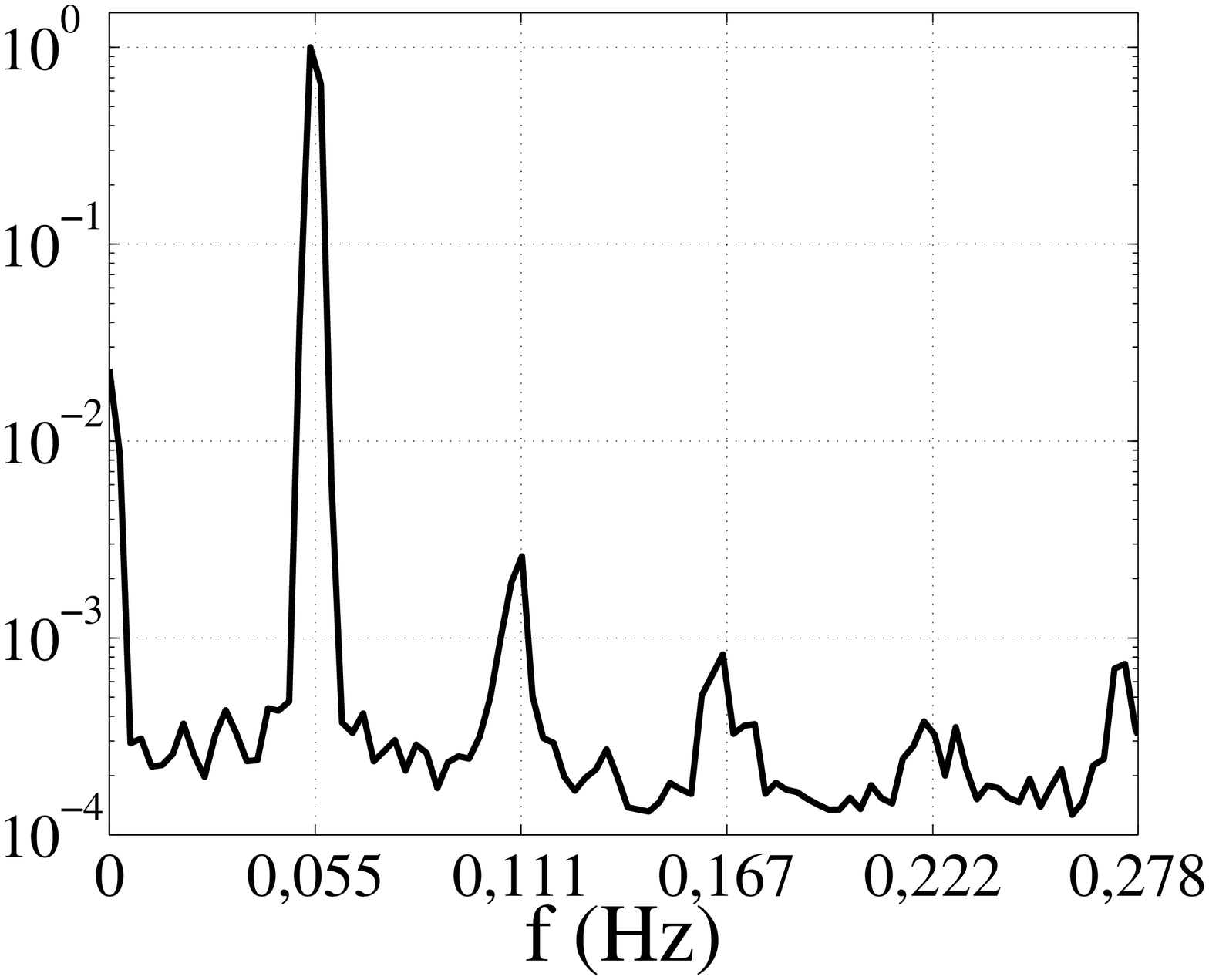}
\end{center}
\caption{(a) A snapshot of the instantaneous velocity field in the absence of topography; the velocity magnitude in cm/s is given by the background color scale and arrows indicate velocity direction. (b) Modal energy content of the horizontal velocity field normalized by the mode-1 amplitude; the inset presents a zoom-in for modes 2 and higher. (c) Normalized, spatially-averaged frequency spectrum of the  horizontal velocity revealing a highly-dominant peak at the fundamental frequency.}\label{FigIncidentMode}
\end{figure}

The first studies investigated the modal content of the internal wave field radiated by the wave generator in the absence of topography, with the goal of establishing the existence of a high-quality, incident mode-1 tide. A snapshot of the velocity field in the vertical mid-plane of the generator ($y=0$) and in the range $x$=3.5 m to $x$ = 5.0 m is presented in figure~\ref{FigIncidentMode}(a). The wave field comprises a sequence of counter-rotating vortices that extend the entire depth of the fluid and are in the process of traveling from left-to-right, consistent with the expected form of a mode-1 wave field in a uniform stratification. It is quite remarkable that even though the internal wave generator only provides horizontal (barotropic) velocity forcing, vertical (baroclinic) velocity disturbances naturally evolve, nicely recreating the velocity distribution of a mode-1 tide. Typical velocities induced are of the order of $10$~mm/s, compared to a forcing amplitude of $A_0\omega=12.2$~mm/s.

A modal decomposition of the wave field presented in figure~\ref{FigIncidentMode}(a) was obtained using a previously developed routine~\cite{Echeverri}, the approach being modified to account for the slight nonlinear features in the stratification. Modal decomposition was performed on data obtained once a time-periodic regime had been clearly established (typically after $15$ periods). A time window of $6$ periods was used for the temporal filtering, and the decomposition was performed at twenty locations within a $0.5$~m wide horizontal window centered around $x=4.5$~m. The modal decomposition was found to be the same at each location, and the experiment was found to be highly repeatable. Further tests were performed, in which the wave field was viewed in the $x-y$ plane from above, to confirm two-dimensionality around the vertical mid-plane $y=0$.

The results of the modal decomposition are presented in figures~\ref{FigIncidentMode}(b) and (c). The nondimensional energy flux associated with the $n^{\mathrm{th}}$ mode is $E_n=A_n^2/n$~\cite{Echeverri}, where $A_n$ is the modal amplitude nondimensionalized by $A_0$. The energy flux in mode-2 is around 2\% of that in mode-1, and the energy flux in the higher modes is barely discernable. The temporal spectrum of the field is so strongly peaked around the forcing frequency that higher harmonics are negligible. Thus, in the absence of topography the experimental arrangement successfully produced a high-quality mode-1 wave field.

The ability to generate a desired vertical modal structure opens the door to systematic experimental studies of internal tide scattering. To investigate this, a Gaussian ridge was carefully placed in the path of the mode-1 wave field, so as not to noticeably disturb the stratification. The maximum topographic slope of the Gaussian was 0.55, corresponding to an angle with respect to the horizontal of $\theta=29^\circ$. The ray slope of the internal tide was 0.67, corresponding to $\theta=34^\circ$. The topography was therefore supercritical, with a criticality of $\varepsilon=1.22$; in this case, significant reflection and scattering of the low-mode internal tide is expected~\cite{Johnston1, Johnston2}.

\begin{figure}
\begin{center}
\includegraphics[width=0.95\linewidth]{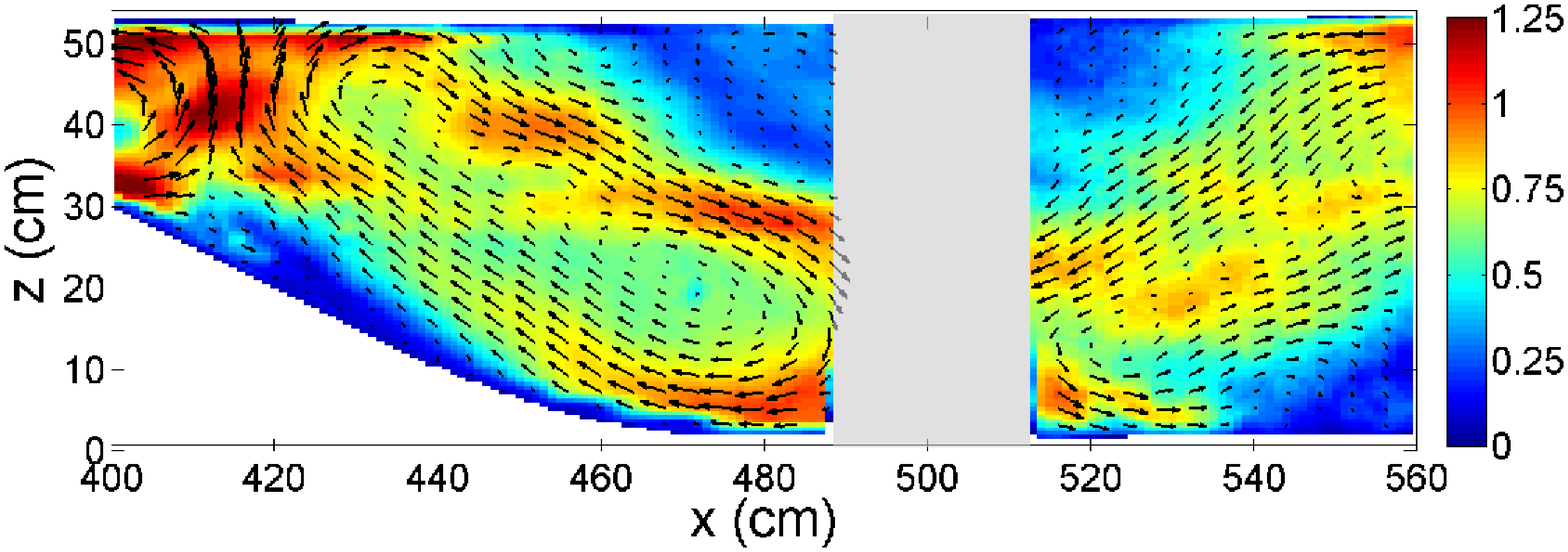}\\
\includegraphics[height=4cm]{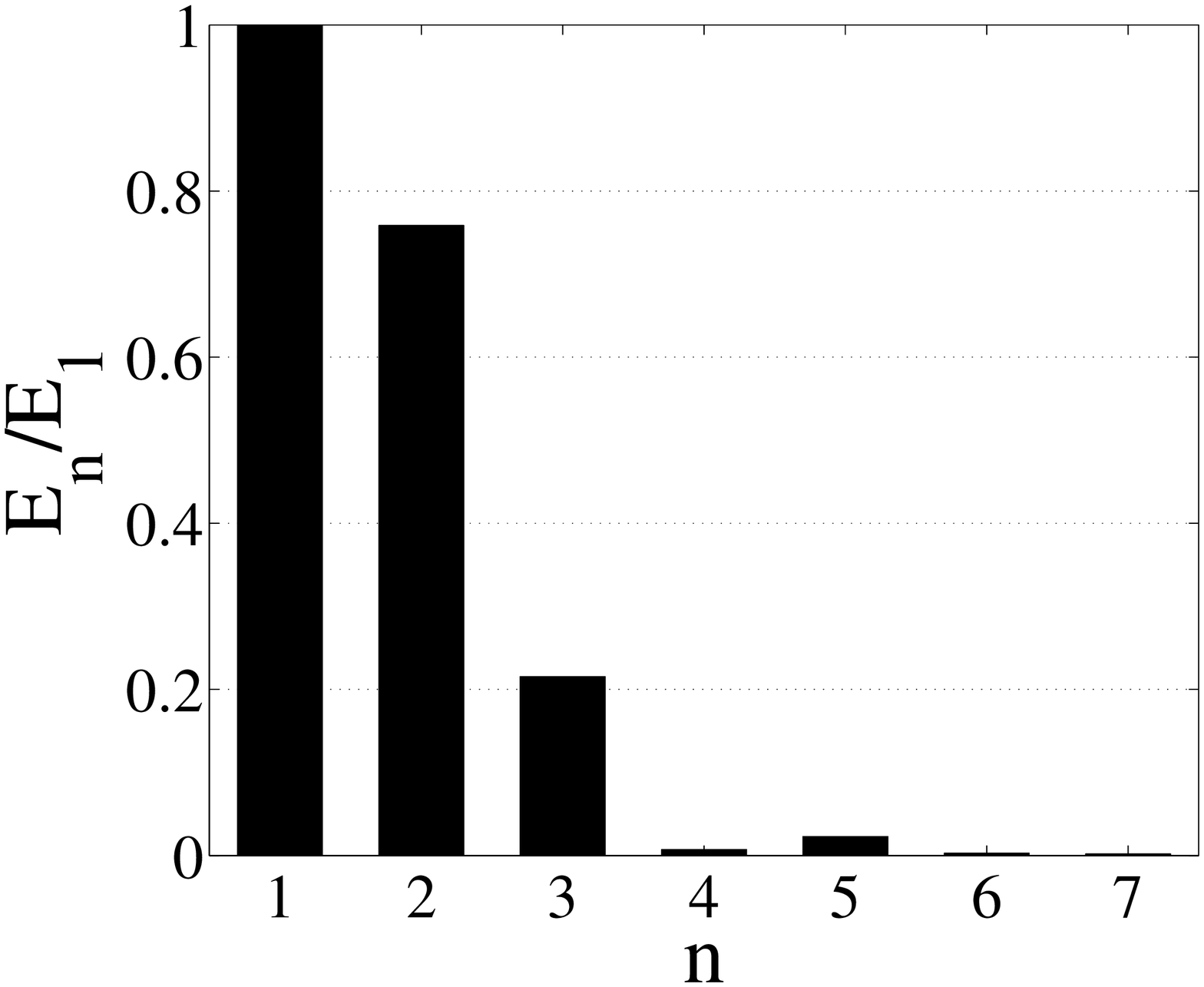}\hspace{2cm} \includegraphics[height=4cm]{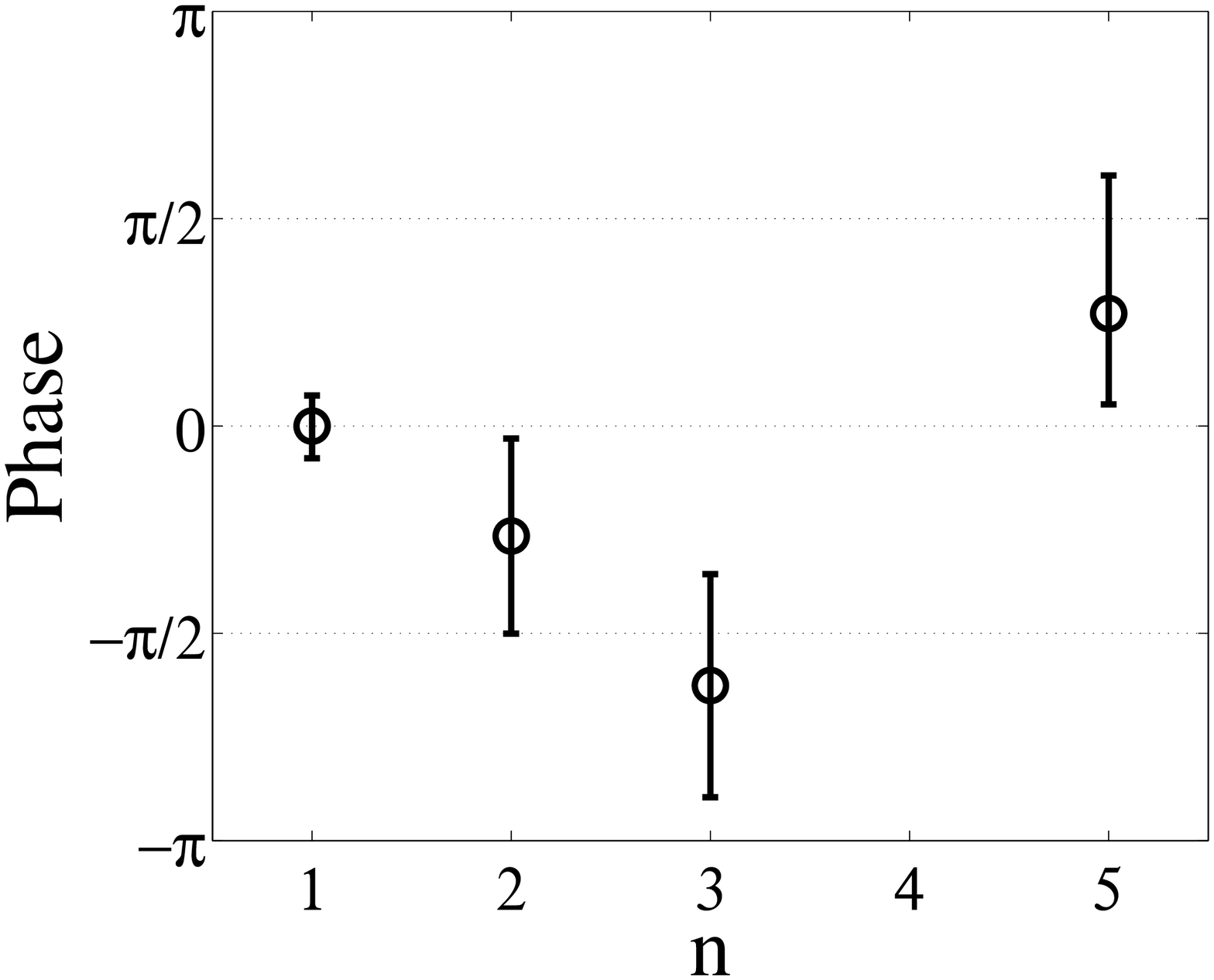}
\end{center}
\caption{(a) Snapshot of the instantaneous velocity field (filtered at the fundamental frequency $\omega$) transmitted by the Gaussian ridge; the velocity magnitude in cm/s is given by the background color scale and arrows indicate velocity direction.
(b) Modal energy content of the horizontal velocity field normalized by the mode-1 amplitude. This data was obtained in the region 5.15 m $<x<$ 5.60 m. (c) Phase of the modes, with the phase of mode-1 arbitrarily being chose to be zero.}\label{FigGaussian}
\end{figure}

A composite snapshot of the wave field transmitted by the Gaussian ridge is presented in figure~\ref{FigGaussian}(a). The snapshot contains a $0.3$~m wide (grey) region with no data, as the second camera was positioned sufficiently far back to ensure modal decomposition could be performed in a region of constant depth. In contrast to the vortex structures evident in figure ~\ref{FigIncidentMode}(a), a clear wave beam structure is visible in figure~\ref{FigGaussian}(a), indicating significant high-mode content of the wave field. The wave beam originates from the top of
the topography, strikes the floor around $x=5.0$ m, where it is reflected upwards. The maximum velocities in the wave beam are roughly $25\%$  higher than in the mode-1 wave field of figure~\ref{FigIncidentMode}(b), suggesting concentration of internal wave energy in the wave beam.

The modal structure of the transmitted wave field, calculated over the range 5.15 m $<x<$ 5.60 m, is presented in figures~\ref{FigGaussian}(b) and (c). The energy flux in mode-2 in the scattered wave field is comparable to that in mode-1, and the energy flux in mode 3 is also significant. Since $E_n=A_n^2/n$, the amplitude of mode-2 is in fact 20\% larger than
that of mode-1, and the amplitude of mode-3 is only 20\% smaller than that of mode-1; it is this rich modal structure that is responsible for the wave beam structure in figure~\ref{FigGaussian}(a). An interesting observation is the existence of a constant phase delay of around $50^\circ$ between modes 1, 2 and 3 (and perhaps even the weak, but detectable, mode 5), as shown by the results presented in figure~\ref{FigGaussian}(c). This suggests the existence of a mechanism for phase locking that arises through the scattering process.

\begin{figure}
\begin{center}
\includegraphics[width=0.95\linewidth]{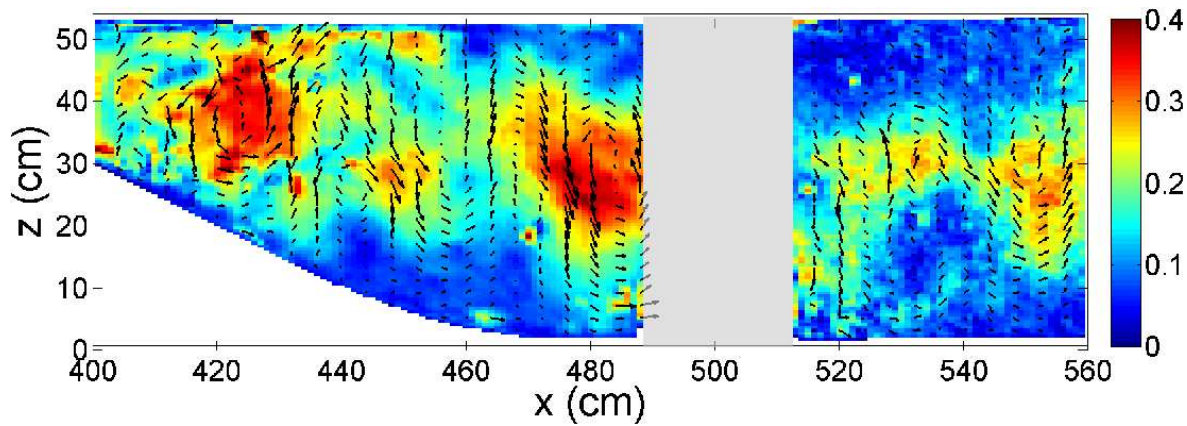}
\end{center}
\caption{The transmitted wave field filtered at the first harmonic frequency $2 \omega$; the velocity magnitude in cm/s is given by the background color scale and arrows indicate velocity direction.}\label{FigSecondHarmonic}
\end{figure}

The amplitude of the transmitted mode-1 signal was $4.3\pm0.1$~mm/s, which is significantly less than incident amplitude of around $10$~mm/s. We furthermore determine that modes 1 through 5 of the transmitted wave field contained only around $40\%$ of the incident energy flux, and thus around $60\%$ of the incident energy was reflected by the topography or transferred to other forms via nonlinear processes. There was clear evidence of the generation of higher harmonics, indicating that nonlinear processes were present. As evidence of this, figure~\ref{FigSecondHarmonic} presents a snapshot of the wave field filtered at the frequency 2$\omega$, revealing first harmonic content within the transmitted wave beam. For this experiment the first harmonic is evanescent, so the observed features must have been locally generated within the wave beam. The major features seem to lie immediately over the topography or near the reflection site, where wave-wave interactions can occur.

In summary, these experiments demonstrate the feasibility to now study internal tide scattering in a laboratory setting, providing a new way to investigate a key candidate processes influencing the fate of oceanic internal tides. These initial experiments provide clear support of predictions that large-amplitude supercritical topography produces significant reflection of the internal tide and transfer of energy from low to high-modes. In addition, we have found some evidence of nonlinear processes that transfer energy to higher harmonics, which may also be significant. It would now seem that a careful comparison between experiment, theory and numerics can be used to resolve any ambiguities regarding the impact of finite-amplitude, rough topography on the low-mode internal tide. This will require some theoretical advancement, however, to handle extended, finite-amplitude, rough topography, perhaps using the Green function approach~\cite{Robinson}.

{\bf Acknowledgments}

We thank J. Sommeria for helpful suggestions and M. Moulin for help
in preparing the cams.
This work has been partially supported by the ANR grant PIWO (ANR-08-BLAN-0113-01)
and the MIT-France program.


\end{document}